\documentclass[conference]{IEEEtran}
\usepackage{cite}
\usepackage{amsmath,amssymb,amsfonts}
\usepackage{algorithmic}
\usepackage{graphicx}
\usepackage{textcomp}
\usepackage{booktabs}
\def\BibTeX{{\rm B\kern-.05em{\sc i\kern-.025em b}\kern-.08em
    T\kern-.1667em\lower.7ex\hbox{E}\kern-.125emX}}
\begin{document}

\title{Ridgeline: A 2D Roofline Model for Distributed Systems}

\author{
\IEEEauthorblockN{Fabio Checconi\textsuperscript{*}}
\IEEEauthorblockA{\textit{Parallel Computing Labs} \\
\textit{Intel}\\
Santa Clara, USA\\
fabio.checconi@intel.com}
\and
\IEEEauthorblockN{Jesmin Jahan Tithi\textsuperscript{*}}
\IEEEauthorblockA{\textit{Parallel Computing Labs} \\
\textit{Intel}\\
Santa Clara, USA \\
jesmin.jahan.tithi@intel.com}
\and
\IEEEauthorblockN{Fabrizio Petrini}
\IEEEauthorblockA{\textit{Parallel Computing Labs} \\
\textit{Intel}\\
Santa Clara, USA\\
fabrizio.fetrini@intel.com}
}

\maketitle

\begingroup
\renewcommand\thefootnote{\fnsymbol{footnote}}
\footnotetext[1]{* Fabio Checconi and Jesmin Jahan Tithi contributed equally to this work and should be cited jointly as co-first authors.}
\endgroup

\begin{abstract}
In this short paper we introduce the Ridgeline model, an extension of the Roofline model~\cite{williams2009Roofline} for distributed systems. The Roofline model targets shared memory systems, bounding the performance of a kernel based on its operational intensity, and the peak compute throughput and memory bandwidth of the execution system.

In a distributed setting, with multiple communicating compute entities, the network must be taken into account to model the system behavior accurately. The Ridgeline aggregates information on compute, memory, and network limits in one 2D plot to show, in an intuitive way, which of the resources is the expected bottleneck. We show the applicability of the Ridgeline on a case study based on a data-parallel Multi-Layer Perceptron (MLP) instance. 
\end{abstract}

\section{Introduction}
The standard Roofline model \cite{williams2009Roofline} ties together a system's floating-point throughput and memory bandwidth, as well as a kernel's operational intensity in a plot that determines visually an upper bound for the kernel's performance on the system being modeled. It is an intuitive way to understand expected peak performance on a shared memory architecture, and it gained widespread adoption in the high-performance computing community. However, the Roofline model is not designed to take into account network bandwidth limitations, which often arise in distributed settings. Recently, the concept of {\it disaggregated system}, that is, a system capable of pooling remote resources and exposing them as local based on application needs~\cite{lim2009disaggregated}, is also getting a lot of attention, and there too the interconnect bandwidth needs to be included in performance modeling.

\begin{figure}[h!]
\centering
\includegraphics[width=\linewidth]{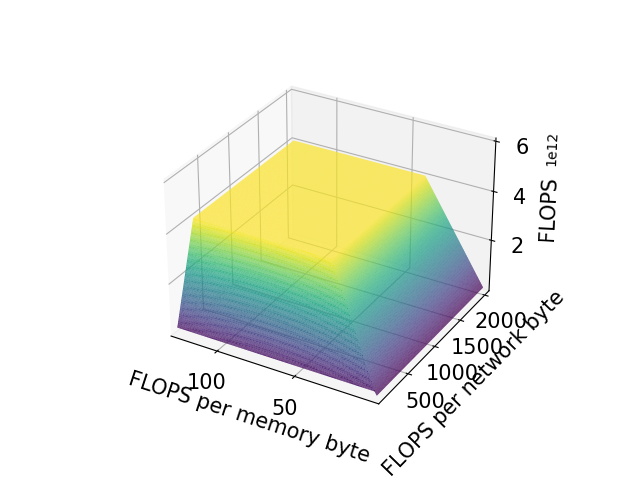}
\caption{A Naive 3D extension of Roofline.}
\label{fig:RR3D}
\end{figure}

One natural extension to the Roofline model for distributed settings can be achieved by generating a 3D plot where the $x$-axis is the operational intensity (expressed in FLOPS per memory byte), the $y$-axis is a measure of communication intensity (expressed in FLOPS per network byte), and the $z$-axis reports the achievable FLOPS. This is the approach followed in \cite{cardwell2019extended}. Figure \ref{fig:RR3D} shows an example of such a 3D model; it can be considered the combination of two separate roofline models, a {\it compute-memory} roofline and a {\it compute-network} roofline, each representing the relation between a pair of resources. We can see that the surface has three faces, each of which corresponds to a different bottleneck. The top plateau corresponds to kernels whose theoretical peak performance is compute bound; the face sloping upwards along increasing operational (resp.\ communication) intensities correspond to memory (resp.\ network) bound kernels.

While this approach is perfectly viable, we are interested in simplifying the creation and interpretation of the model, and find a meaningful projection on a plane, trying to bring back the direct and intuitive nature of the original Roofline.

To that purpose, we introduce the {\bf{Ridgeline}} model, combining the roofline plots for multiple resources into a planar plot, and allowing the immediate visual identification of the bottlenecks for peak kernel or application performance. In the rest of the paper we go into the details of the model, and we provide a case study to illustrate some of its uses.

\section{The Ridgeline Model}
The Ridgeline model is based on the properties of the plane defined by the intensity measures used in Roofline analysis, and how they relate to peak-theoretical performance for any given architecture. Table~\ref{tab:Symbols} lists the symbols we use for the various quantities at play.

We consider the plane defined by arithmetic and  memory intensity, though it would be possible to carry out an equivalent analysis using arithmetic and network intensity. For the scope of this paper, we define memory intensity as the bytes of memory accessed per byte of network traffic, and network intensity as the FLOPS per byte of network traffic.

\begin{table}[htbp]
  \centering
  \caption{Symbols and Their Meanings}
 
    \begin{tabular}{lp{0.5\linewidth}}\toprule
    Symbol & Meaning \\\midrule
    $F$     & FLOPS per work unit \\
    $B_M$   & Memory bytes accessed per work unit \\
    $B_N$   & Bytes transferred over the network per work unit \\
    \(I_A\) & Operational (Arithmetic) Intensity, $I_A=F/B_M$  \\
    \(I_M\) & Memory Intensity, $I_M=B_M/B_N$  \\
    \(I_N\) & Network Intensity, $I_N=F/B_N$  \\
    \bottomrule
   \end{tabular}
  \label{tab:Symbols}%
\end{table}%

\begin{figure*}[tbph]
\centering
\includegraphics[width=\linewidth]{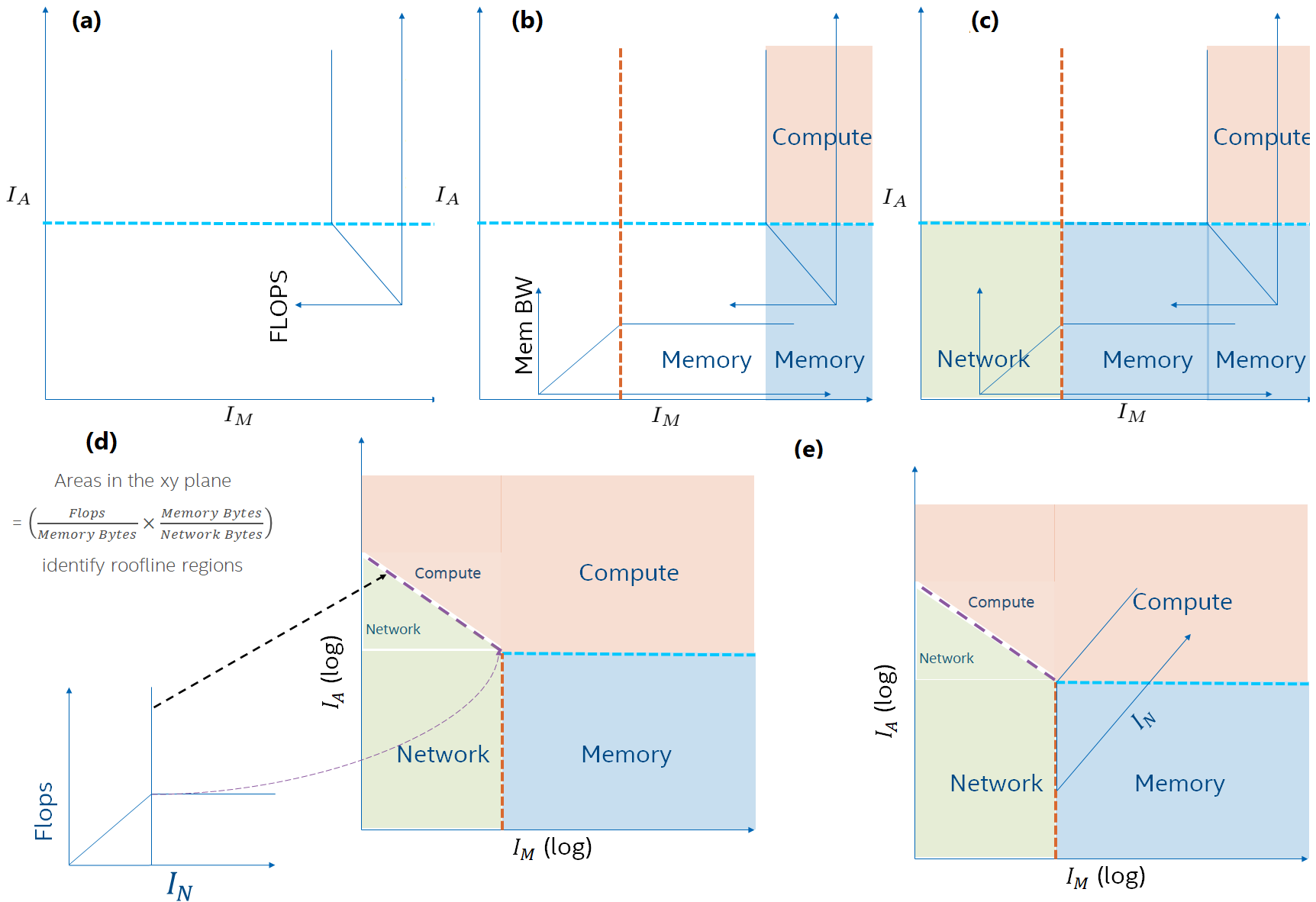}
\caption{Illustration of the Ridgeline principles.}
\label{fig:RR3}
\end{figure*}

Consider Figure \ref{fig:RR3}a, showing the plane defined by memory and arithmetic intensity. Since the vertical axis is the same as the $x$ axis a traditional Roofline uses, it is possible to superimpose a Roofline whose FLOPS axis is perpendicular to the plane of the intensities. The blue dashed line passing through the knee of the Roofline curve divides the plane into two regions. Based only on arithmetic intensity we can say that the region above the line is compute bound, and the region below is memory bound.

To complete the analysis, including the effect of network bandwidth, consider Figure \ref{fig:RR3}b. Here the memory-network roofline is superimposed along the shared $x$ axis. While this form of roofline is not commonly used, it is based on the same principles as its more frequently encountered relatives, expressing the memory bandwidth achievable as a function of memory intensity. As the traditional Roofline, for low values of the intensity the throughput measure (memory bandwidth in this case) grows linearly with the first limiting factor (network bandwidth), until the intensity is high enough to saturate the memory bandwidth of the system.
The red dashed line, corresponding to the balance point in the roofline between memory and network bandwidth, divides the plane into two regions, the one to its left network bound, the one to its right memory bound.

By considering the four quadrants defined by the two dashed lines and combining the bottlenecks identified by the two models, we can derive the conclusions that follow, also shown in Figure~\ref{fig:RR3}c. In the lower-left quadrant the limiting factor is the network bandwidth, with the traditional Roofline indicating memory as the bottleneck, and the memory-network roofline asserting the dominance of the network over the memory. Based on the same observation, but the different side in the memory-network Roofline, in the lower-right quadrant the limiting factor is the memory. Applying the same reasoning we can also conclude that the upper-right quadrant is compute-bound.

In the upper-left quadrant we need an additional step to be able to determine the limiting factor. From the two roofline models we know that we can be limited either by network bandwidth or compute, but we have to establish a relationship between these two quantities to find which imposes the lowest bound.

We start by observing that the product of the coordinates $xy=\mathrm{FLOPS}/B_C$, so for any value of $k$, the performance at the points for which $xy=k$ is satisfied only depends on network intensity. In Figure \ref{fig:RR3}d, we see that for increasing values of $k$, the points in the plane move progressively towards being compute bound. The separation line between the network and compute bound regions is obtained when $k$ is the balance point between the system compute throughput and network bandwidth. Note that the separation is on a straight line because of the logarithmic scale on both axes.

Finally, Figure~\ref{fig:RR3}e, shows the regions in the plane and the relative bottlenecks.

\begin{figure}[tbph]
\centering
\includegraphics[width=0.9\columnwidth]{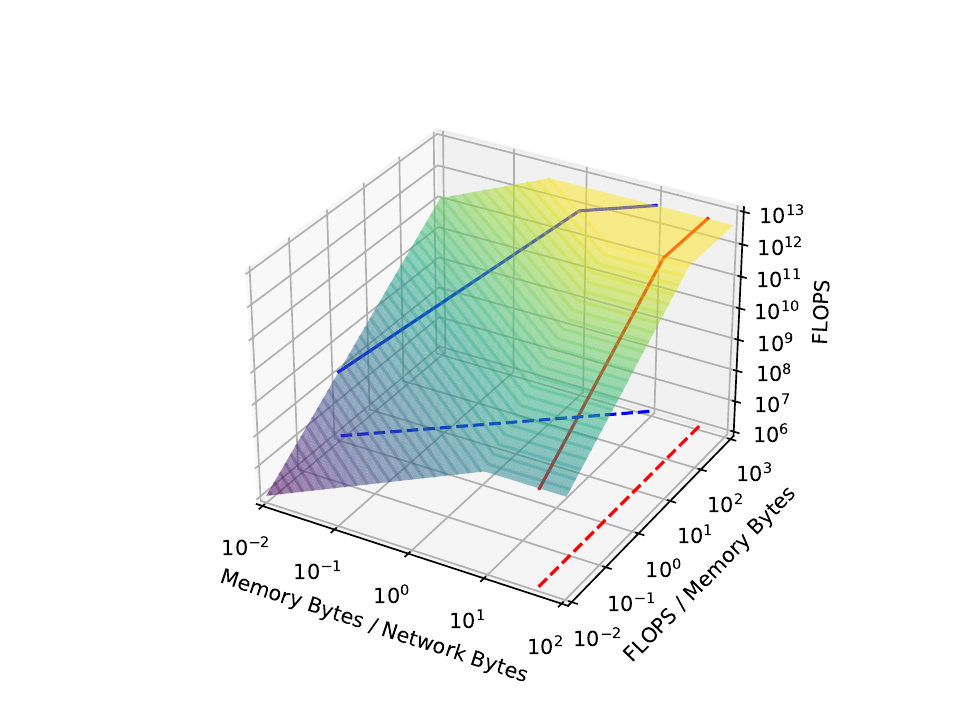}
\caption{The surface showing FLOPS as a function of network and memory intensities.}
\label{fig:proj-3d}
\end{figure}

Figure \ref{fig:proj-3d} shows an alternative way of visualizing FLOPS as a function of network and memory intensities. The red line in figure corresponds to a traditional Roofline plot for a given value of network intensity, shown by the dashed red line. The blue line corresponds to a compute-network roofline for a given value of arithmetic intensity, and its projection is shown in the dashed blue line. This derives from the fact that $I_N=xy$, and the compute-network roofline is plotted by varying $t$ in $xy=t$.

\section{Case Study}

To illustrate the benefits of a model more expressive than the traditional Roofline, we consider the multi-layer perceptron (MLP), a key component of several machine learning models. In particular, we are interested in models that include MLPs and are deployed on a distributed computation environment. As is the case with most machine learning kernels, the behavior of MLPs depends on the parameters used in each specific application, and the one we consider here is Facebook's DLRM recommendation model~\cite{mudigere2021high}. 

\begin{figure*}[hpbt]
\centering
\includegraphics[width=\linewidth]{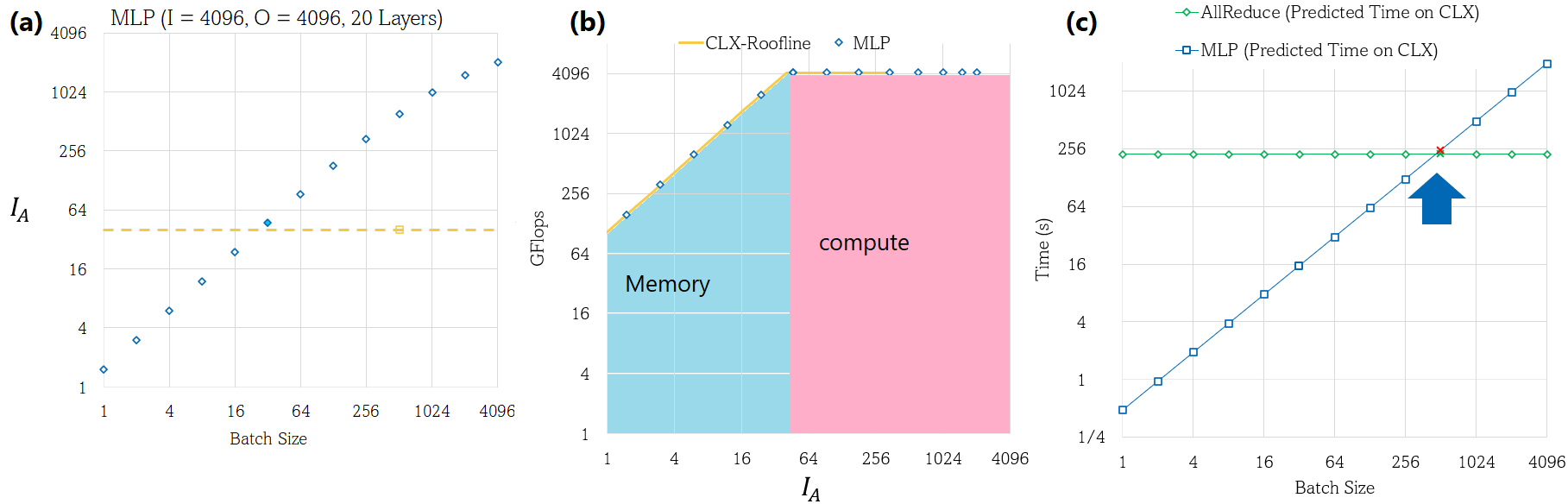}
\caption{Roofline for an MLP instance on a Cascade Lake system. Analysis done on multiple points, corresponding to different batch sizes.}
\label{fig:RR1}
\end{figure*}

\begin{figure}[htbp]
\centering
\includegraphics[width=0.4\linewidth]{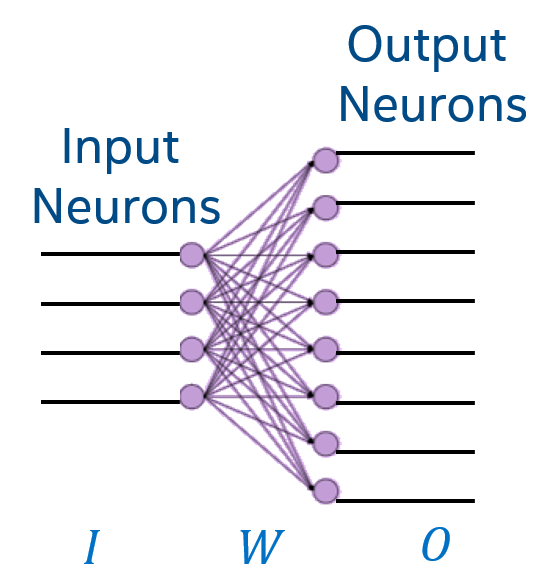}
\caption{Individual layer of a Multi-layer Perception Network.}
\label{fig:MLP}
\end{figure}

MLPs are composed of a sequence of fully connected layers, also known as linear layers, in which every input neuron is connected to every other output neuron (see Figure \ref{fig:MLP}), and are commonly used as stages in various neural network topologies. MLP neurons have the following input-output relationship: \(O_{l} = f(W_l \times I_l+b_l)\), where $l$ is the layer id, \(O_l\) is the output of layer $l$, \(I_l\) is the input, $W_l$ are the weights and \(b_l\) is the bias. 
The matrices \(\{W_i\}\) and the vectors \(\{B_i\}\) are parameters that need to be trained. 

\begin{table}[htbp]
  \centering
  \caption{MLP Parameters for a given layer}
    \begin{tabular}{lp{.7\linewidth}}\toprule
    Parameter & Size  \\\midrule
    O   & batch size $\times$ output feature map size \\
    I   & batch size $\times$ input feature map size \\
    W   & input feature map size $\times$ output feature map size \\
    b   & output feature map size \\
    \bottomrule
    \end{tabular}%
  \label{tab:MLP}%
\end{table}%
Three performance of a fully connected layer can be modeled as a function of three parameters: batch size, input feature map size, and output feature map size. Table \ref{tab:MLP} describes the sizes of each variable for a given layer.
There are three computation phases in the deployment of an MLP: forward propagation, activation gradient computation, and weight gradient computation. All of these can be directly expressed as dense matrix-matrix multiplications (GEMMs). Other low-order components are matrix transpositions, updates of weights and biases, and sum of biases to the products which we assume can be ignored during a Roofline analysis.

If we consider a data-parallel implementation for MLP training, a batch of input is split into mini-batches and those mini-batches are distributed among the compute nodes. In addition, each compute node will have its local copy of the weights and biases which will be trained/optimized locally with the mini-batches being processed by the compute node. At the end of a given epoch (once all inputs are processed from all mini-batches), the weights and biases are synchronized across all compute nodes using some type of all-reduce operation that would use the network/interconnect bandwidth, in a distributed setting. This is the use case that we consider in this section. 

\begin{figure*}[tbph]
\centering
\includegraphics[width=\linewidth]{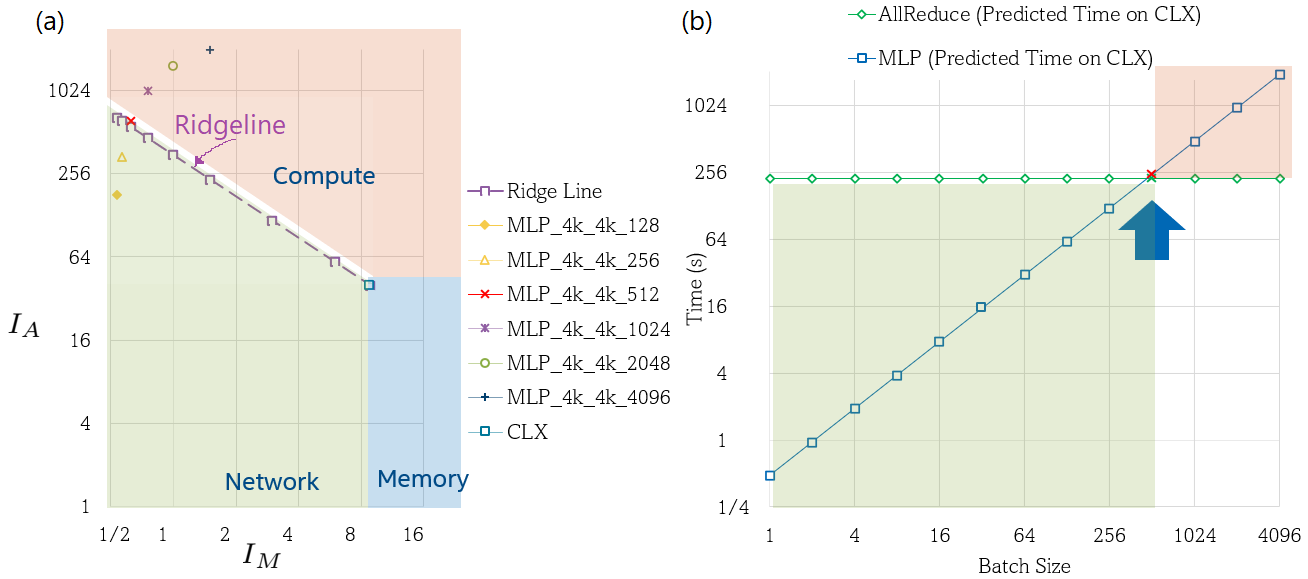}
\caption{Ridgline for a MLP instance, showing how the Ridgeline regions map to the projected performance.}
\label{fig:RR2}
\end{figure*}

Figure \ref{fig:RR1}a shows the arithmetic intensity of MLPs for different batch sizes when input output feature map size of \(4096\). The figure shows that as batch size increases, the arithmetic intensity also increases because one of the three dimensions of the underlying matrix multiplication increases. If we consider strong scaling, in the case of data parallelism, the batch size would decrease as we scale the number of compute nodes. One would like to strong scale usually for two reasons: 1) to speed up the training process, 2) to have a batch size that fits in a given node's memory. Therefore, in the case of strong scaling, the arithmetic intensity will decrease with an increasing number of nodes, and whether an MLP computation would be computed bound or memory bound would depend on the supported operational intensity of the given hardware the MLP kernel is running on. For example, in Figure \ref{fig:RR1}a the yellow horizontal line shows the arithmetic intensity of a state-of-the-art Intel Xeon Cascade Lake (CLX) node with 4.2TF/s FP32 flop capacity, 105 GB/s memory bandwidth, and 12 GB/s network bandwidth per socket. The MLPs with arithmetic intensity higher than the yellow line (batch size 32 or higher) have the potential to reach the peak FLOPS of the platform; these would be ideally compute-bound and would be located on the horizontal portion of the standard Roofline model as shown by Figure~\ref{fig:RR1}b. MLPs with a batch size smaller than 32 would be limited by memory bandwidth according to the standard Roofline. However, this Roofline does not consider the cost for all-reduce that would need to happen after the computation of the MLP is done to synchronize the weights and the biases. Figure \ref{fig:RR1}c shows that in theory, on this CLX machine, up to batch size 512, it would take more time to do the all-reduce than to do the actual MLP computation. The standard Roofline fails to capture this information.

Figure \ref{fig:RR2}a shows the Ridgeline plot for the MLPs starting from batch size 256. In a Ridgeline plot, the $x$-axis shows the memory bytes per network byte and $y$-axis shows the FLOPS per memory bytes. The central point of the ridgeline is located at x=\(Memory ~ BW/Network ~ BW\), y=\(Flops/ Memory ~BW\) for the given architecture (CLX in this case). The 2D Ridgeline separates out the three regions bounded by compute, memory, and network clearly and on this plot, MLP with batch size 512 is indeed on the ridgeline. Thus, with the Ridgeline plot we can see that MLPs with batch size 1024 and higher would be compute-bound and any batch size lower than 512 would be network bound on the CLX nodes we are considering. The bottleneck assessment from the ridgeline plot is confirmed by the timing projections shown in Figure~\ref{fig:RR2}b.

Based on this bounding region identified by the Ridgeline, it is easy to compute the expected peak performance. If the bounding region is memory, the runtime can be found by dividing the memory traffic by the memory bandwidth. If the bounding region is the network, the runtime can be found by dividing the network traffic (all-reduce volume in this case) by the network bandwidth and if the bounding resource is compute, the runtime can be found by diving the total flops by the flop capacity of the hardware. Also, note that all points on the Ridgeline would produce the same GFLOPS/s because on that line $xy=k$: the product of $x$  and $y$-axis value is a constant and that can be found by multiplying \(Flops/Network~Bytes\) of the kernel by the network bandwidth of the system.

\section{Conclusion}
We presented a 2D Ridgeline model which can be used to identify bounding system resources while running an application or a kernel on a distributed system. We have used the Ridgeline model to analyze the performance of Facebook's DLRM model and we plan to publish that analysis result in a full paper in the future.
%\newpage
\bibliographystyle{plain}
\bibliography{references}
\end{document}